\documentclass[runningheads]{llncs}
\usepackage[T1]{fontenc}
\usepackage{graphicx}
\usepackage{amsmath}
\usepackage{amsfonts}
\usepackage{amssymb}
\usepackage{multirow}

\usepackage[
backend=biber,
style=numeric,
sorting=none
]{biblatex}
\usepackage{xcolor}
\graphicspath{{./Figs/}}
\begin{document}
\title{Control of Cellular Automata by Moving Agents with Reinforcement Learning}
\titlerunning{Control of CA by Moving Learning Agents }
\author{Franco Bagnoli\inst{1,2,3\dagger}\orcidID{0000-0002-6293-0305} \and
Bassem Sellami\inst{2}\orcidID{0000-0001-6869-3518} \and
Amira Mouakher\inst{2}\orcidID{0000-0002-1346-3851}\and
Samira El Yacoubi\inst{2}\orcidID{0000-0002-8017-5286}
}
\authorrunning{F. Bagnoli et al.}

\institute{Department of Physics and Astronomy and CSDC, University of Florence,  via G. Sansone 1, 50019 Sesto Fiorentino (Italy) \email{franco.bagnoli@unifi.it}, 
\and Espace-Dev, UPVD, IRD, UM, Perpignan (France) \email{\{bassem.sellami, amira.mouakher,yacoubi\}@univ-perp.fr}
\and INFN, Sect. Florence (Italy) 
}
\maketitle             

\let\at@
\catcode`@=\active
\def@#1{\ifmmode\boldsymbol{#1}\else\at#1\fi}

\let\quot"
\catcode`"=\active
\def"#1"{``#1''}

\newcommand{\eq}[2][]{%
    \ifthenelse{\equal{#1}{}}{%
        \begin{equation*}
            #2%
        \end{equation*}%
    }{%
        \begin{equation}\label{eq:#1}%
            #2%
        \end{equation}%
    }%
}
\newcommand{\meq}[2][]{%
    \ifthenelse{\equal{#1}{}}{%
        \begin{equation*}%
            \begin{split}%
                #2%
            \end{split}%
        \end{equation*}%
    }{%
        \begin{equation}\label{eq:#1}%
            \begin{split}%
                #2%
            \end{split}%
        \end{equation}%
    }%
}

\newcommand{\eps}{\varepsilon}

\newcommand{\Eq}[1]{Eq.~\eqref{eq:#1}}

\begin{abstract}
In this exploratory paper we introduce the problem of  cognitive agents that learn how to modify their environment according to local sensing to reach a global goal. We concentrate on discrete dynamics (cellular automata) on a two-dimensional system. We show that agents may learn how to approximate their goal when the environment is passive, while this task becomes impossible if the environment follows an active dynamics.

\end{abstract}
\section{Introduction}
In this paper we investigate the problem of cognitive agents that learn how to modify their environment, in particular, a two-dimensional Boolean Cellular Automaton (CA) system, in order to achieve a certain goal (the asymptotic density of ``one'' cells). 

The environment is modeled as a parallel outer totalistic CA. The environment evolves while agents learn or apply their learned behavior. As we shall see, the most modifiable environment is represented by the identity rule, while other CA rules may represent an impossible obstacle. 

Agents are represented as totalistic, probabilistic cellular automata with a sensing area (the Moore neighborhood) and an actuator ares (the central cell). Each agent has a global goal: to learn a probabilistic, totalistic rule so that, when applied, it may brings the average density of the environment towards a given target.  Agents ``learn'' by modifying their transition probabilities, which, after many trials, become in general deterministic (either zero or one). 

The difficulty of the task is mainly due by the environment evolution, i.e., the ``physical'' world. In the case in which this evolution is passive (all modifications are  maintained, identity rule), the control is easy and agents quickly learn the corresponding probabilities, which, when applied, lead the system to a state which  approximates the target. 

In the case of more complex environments, which evolve following an active dynamics, learning can be hindered by the lack of examples. In this case may be of help the presence of other agents, which vary the environment. In any case, in active environments, the action of agents generally fails to achieve the desired result, and correspond to small modifications of the ``natural'' asymptotic density. 

The outline of this paper is the following. The model is defined in Section~\ref{sec:model}. Since the action of agents is to learn a totalistic rule, Section~\ref{sec:TCA} is devoted to the study of the possible patterns and asymptotic densities of such a class of automata. Section~\ref{sec:learning} describes the learning process. Section~\ref{sec:identity} presents the results in the case of a passive environment, while Section~\ref{sec:complex} discusses more complex cases. Conclusions are drawn in the last section.

\section{The model}\label{sec:model}

The model is composed by two main elements: 
\begin{enumerate}
    \item The physical word, here represented by a two-dimensional outer totalistic Boolean CA.
    \item The moving agents, that can perform measurements on a given ``sensing'' area and modify an ``actuator'' area, according with a ``strategy'' (a probabilistic totalistic rule) for reaching a target (a global density of ``one'' cells) that is the subject of learning. 
\end{enumerate}

In this work we examine the simple case of agents whose sensing area is the set of 9 cells, i.e., the Moore neighborhood (all cells with a distance less or equal to $\sqrt{2}$ from the central cell), and the only measured quantity is the number of ``one'' cells. The actuator area is just the central cell. 
\begin{figure}
    \centering
    \includegraphics[width=0.5\linewidth]{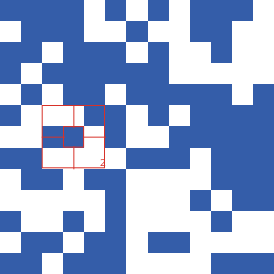}
    \caption{An agent (red lines) in its environment. The larger square denotes the sensing ares, the smaller square is the  actuator area, the number on the lower-right corner is the target number $\bar{m}$ of ``one'' cells (in figure the measured number $m$ is three).}
    \label{fig:model}
\end{figure}

Let us indicate with $M$ the number of cells in the sensing area (see Fig.~\ref{fig:model}). The measurement quantity is therefore a number $0\le m\le M$.  The control strategy is a vector $P(m) \equiv P(1|m)$ which gives the probability of imposing a value 1 on the target cell, given a measurement $m$. Agents have a goal, that for simplicity can be taken as reaching a  target number of one cells in the sensing area $\bar{m}$, which in principle should correspond to a density $\bar{\rho} = \bar{m}/M$. 

The challenge for the learning agent is that of elaborating a strategy (an updating scheme of the actuator area based on the values in the sensing one) able to drive the system towards the desired density. 

Notice that, while the updating of the ``world'' is parallel (all cells update their value at the same time, based on the state of their own neighborhood), the action of one agent is serial, on the actuator cell. Therefore, the problem for the agent can also be reformulated as \emph{find the serial (or asynchronous) updating function that, combined with the ``natural'' parallel updating of the system, can drive it to a desired density}. 

The action of more than one agent is that of partial asynchronism combined with the parallel updating. 
We shall limit our present investigation to Boolean functions that depend on the Moore neighborhood, for which $M=9$. 

Before dealing with the learning procedure, let us introduce the dynamics of asynchronous totalistic cellular automata, and the possible outcome of a ``pure'' (i.e. a deterministic) strategy, using the language of game theory. 

\section{Totalistic and outer totalistic Boolean cellular automata}\label{sec:TCA}

\begin{figure}
    \centering
    \begin{tabular}{cc}
     $MGE4$ & $MGE6$ \\
     {\includegraphics[width=0.4\linewidth]{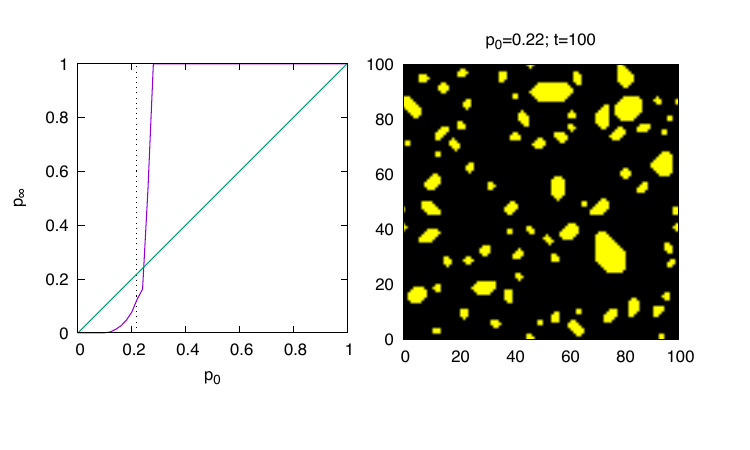}} & 
    {\includegraphics[width=0.4\linewidth]{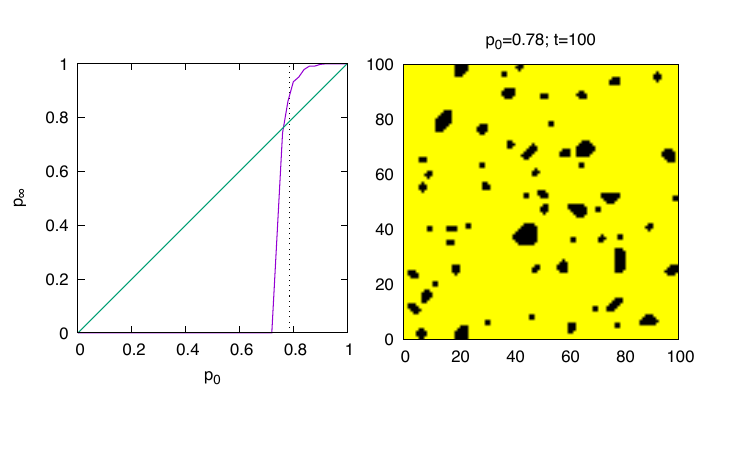}} \\

    $MGE5$ &  $MGE5$ \\
    {\includegraphics[width=0.4\linewidth]{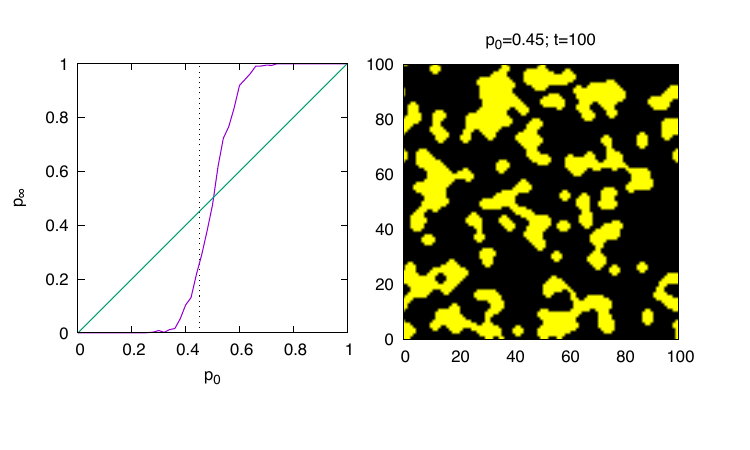}} &
    {\includegraphics[width=0.4\linewidth]{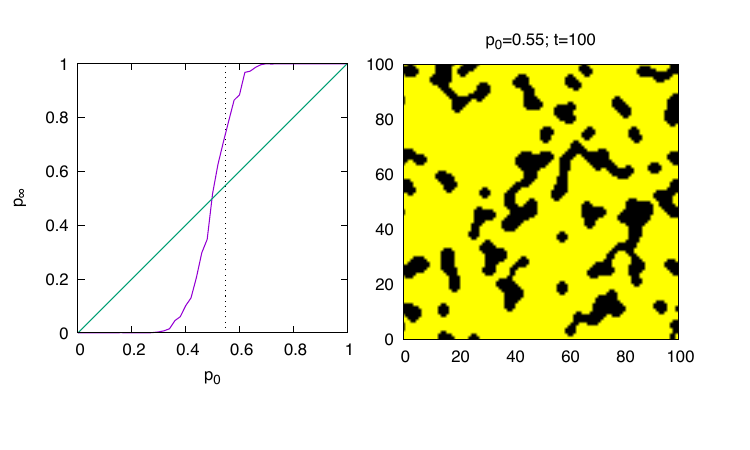}} \\
  \end{tabular}
    \caption{Asymptotic patterns ($100\times100$ sites) of parallel updating of totalistic majority  rules $MGEX$, $T=100$. The asymptotic patterns and densities depends on the initial density, except for $X\le 3$ (all ones) and $X\ge 7$ (all zeros). The transitions for $X=4$ and $X=6$ are dominated by nucleation, that for $X=5$ shows a gradual transition of cluster sizes (spinodal decomposition).}
    \label{fig:GE}
\end{figure}

\begin{figure}
    \centering
    \begin{tabular}{ccc}
     $MLE0p$ & $MLE1p$ & $MLE2p$\\
     {\includegraphics[width=0.33\linewidth]{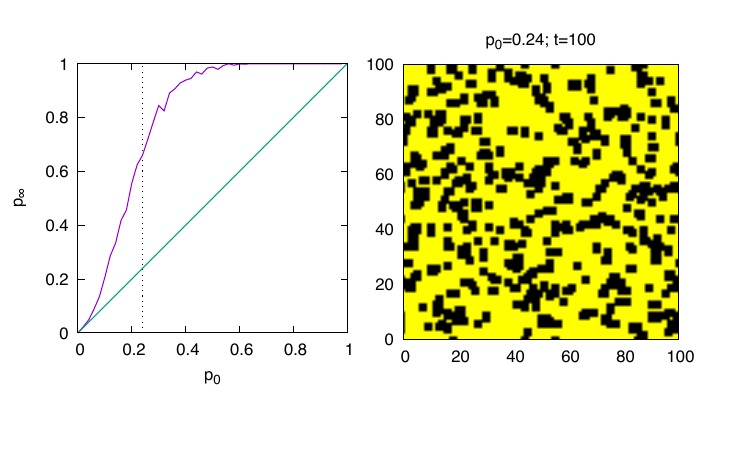}} & 
    {\includegraphics[width=0.33\linewidth]{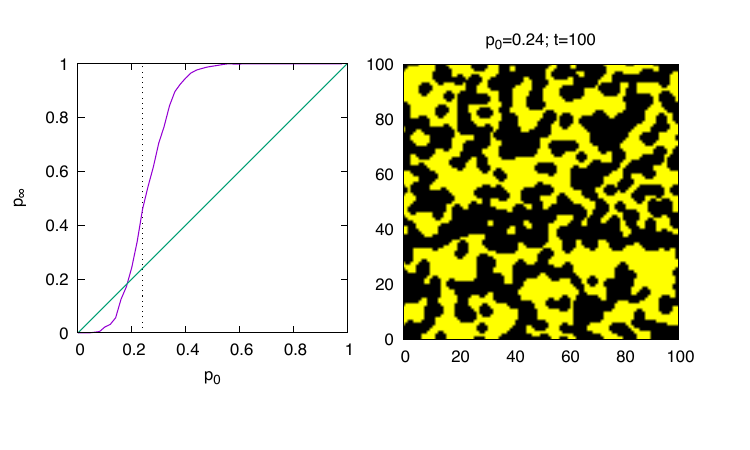}} &
    {\includegraphics[width=0.33\linewidth]{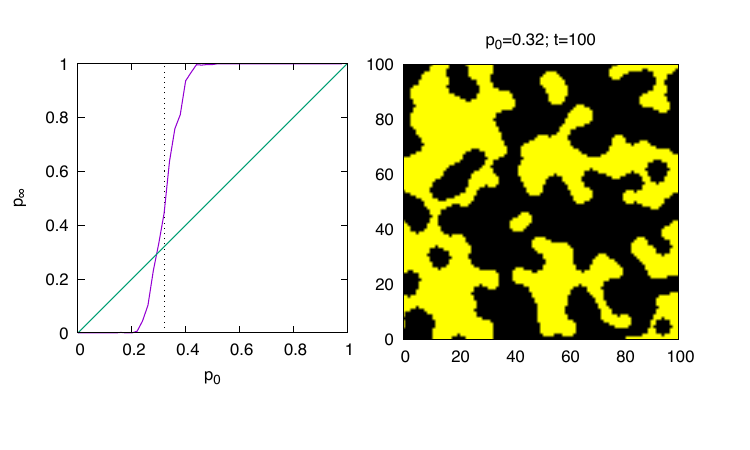}} \\

    $MLE3p$ &  $MLE4p$ &  $MLE5p$ \\ 
    {\includegraphics[width=0.33\linewidth]{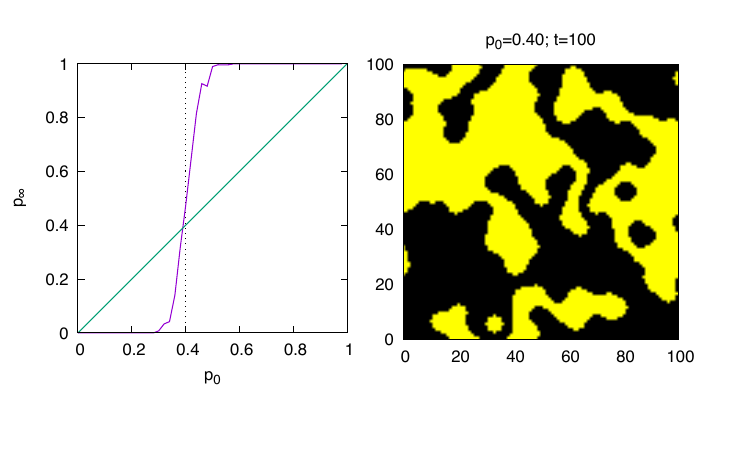}} &

    {\includegraphics[width=0.33\linewidth]{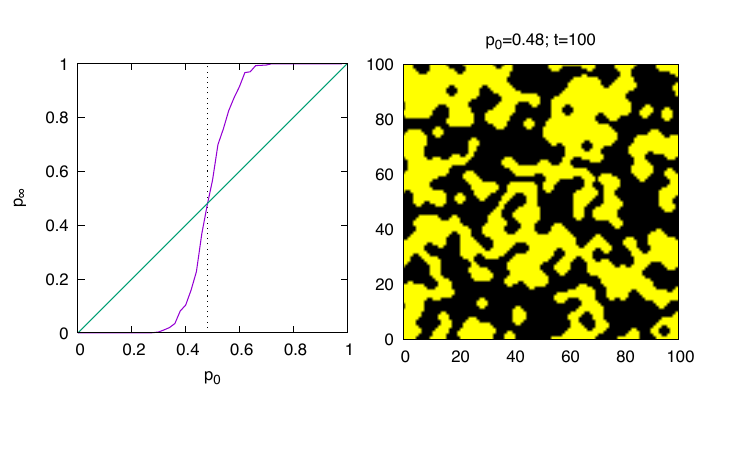}} &
    {\includegraphics[width=0.33\linewidth]{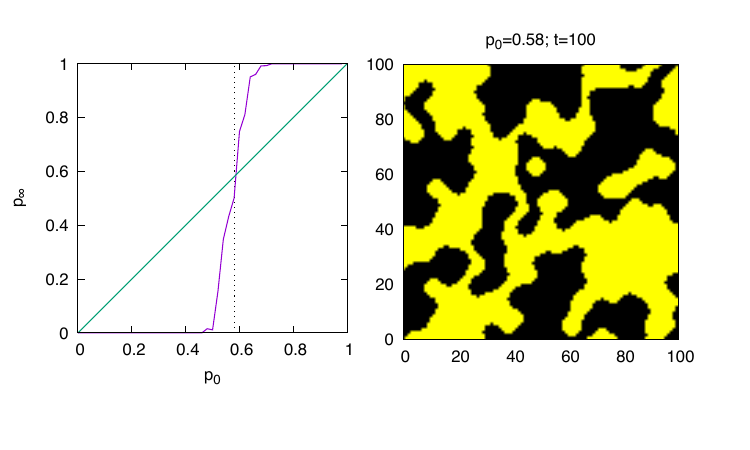}} \\
    $MLE6p$ &  $MLE7p$ & $MLE8p$ \\
    {\includegraphics[width=0.33\linewidth]{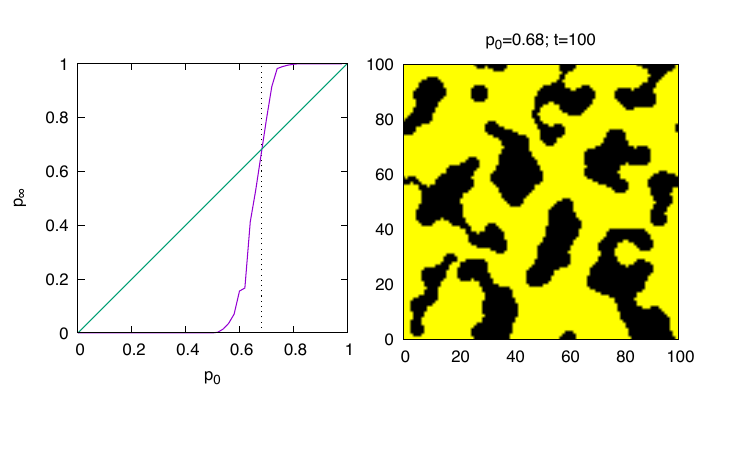}} &
    {\includegraphics[width=0.33\linewidth]{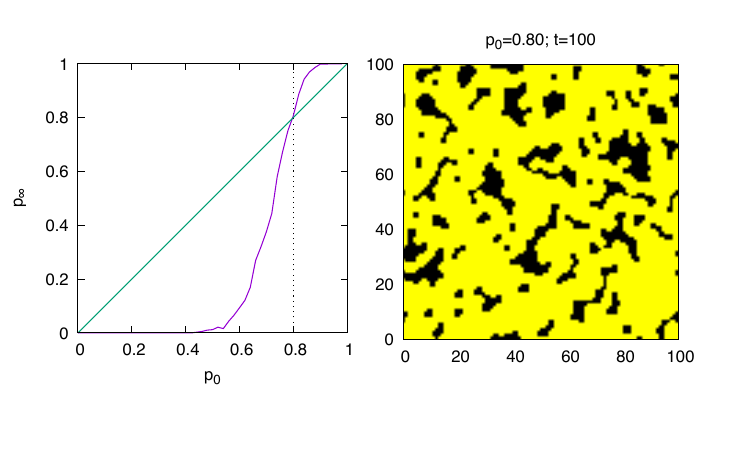}}   & 
    {\includegraphics[width=0.33\linewidth]{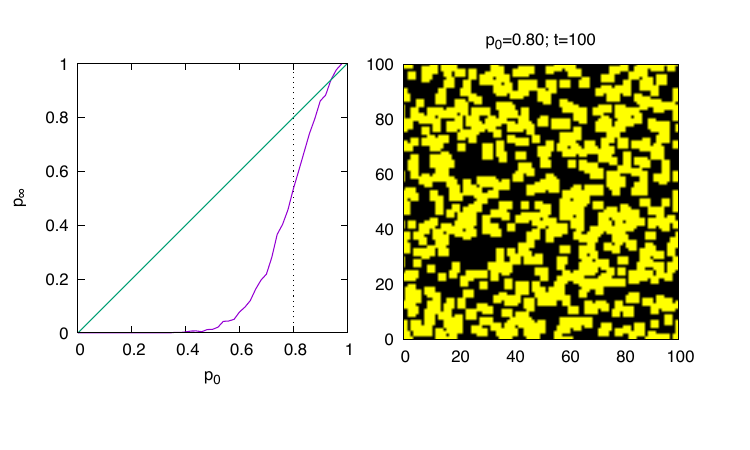}} \\
  \end{tabular}
    \caption{Results for lattices of $100\times100$ sites of fully synchronous updating of totalistic minority  rules $MLEXp$,  $T=100$. 
    Left: asymptotic density $\rho_\infty$ vs initial density $\rho_0$. Right: a snapshot of the configuration for $\rho_0$ corresponding to the red dotted line at left.
    The asymptotic patterns and densities depends on the initial density, except for $X=9$ (all ones), $T=100$. Patterns and curves take opposite values for odd time steps.}
    \label{fig:LEp}
\end{figure}

\begin{figure}
    \centering
    \begin{tabular}{ccc}
     $M\!LE0s:\rho_\infty=0.15$ & $M\!LE1s:\rho_\infty=0.25$ & $M\!LE2s:\rho_\infty=0.35$ \\
     {\includegraphics[width=0.33\linewidth]{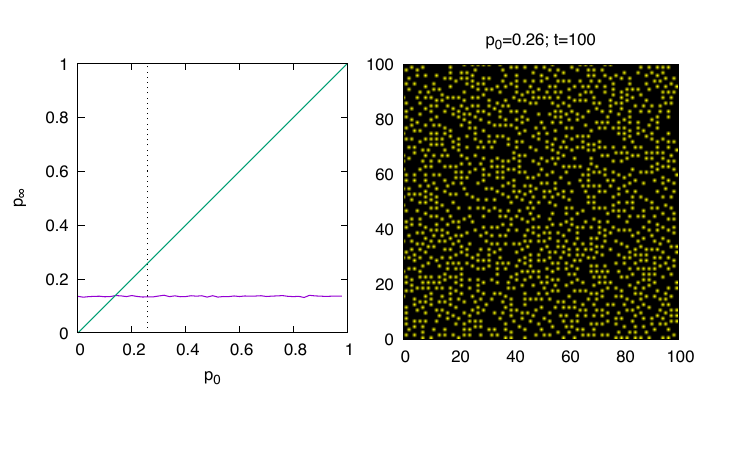}} & 
    {\includegraphics[width=0.33\linewidth]{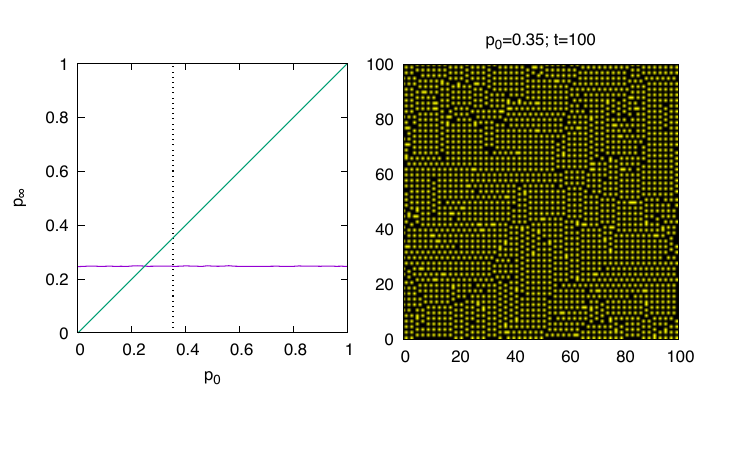}} & 
    {\includegraphics[width=0.33\linewidth]{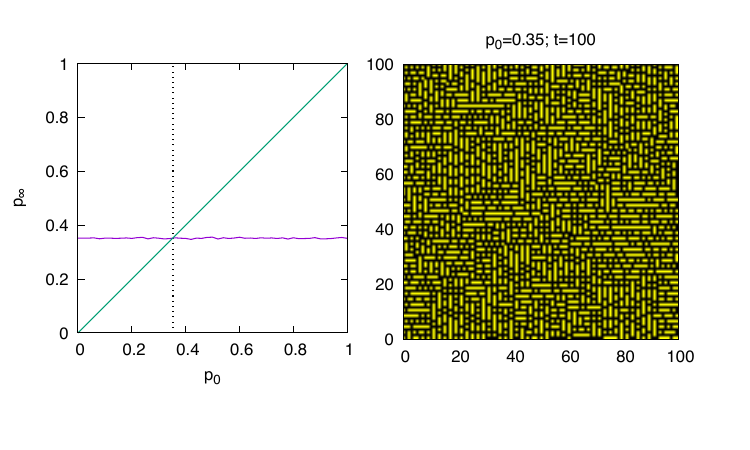}} \\
    
    $M\!LE3s:\rho_\infty=0.5$ &  $M\!LE4s:\rho_\infty=0.5$ &  $M\!LE5s:\rho_\infty=0.5$ \\
    {\includegraphics[width=0.33\linewidth]{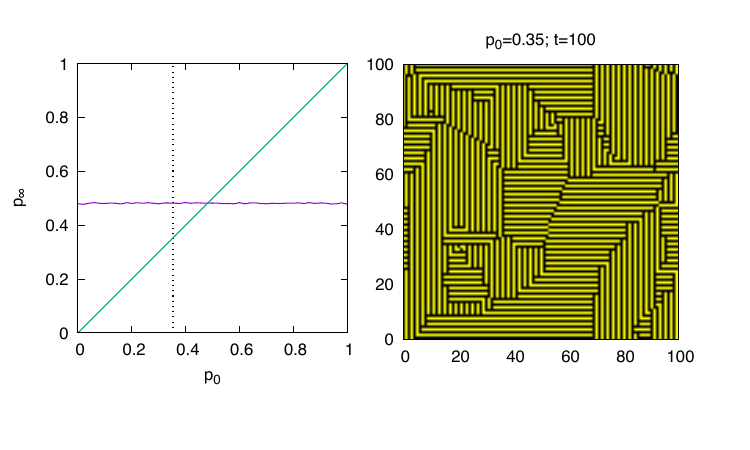}} &
    {\includegraphics[width=0.33\linewidth]{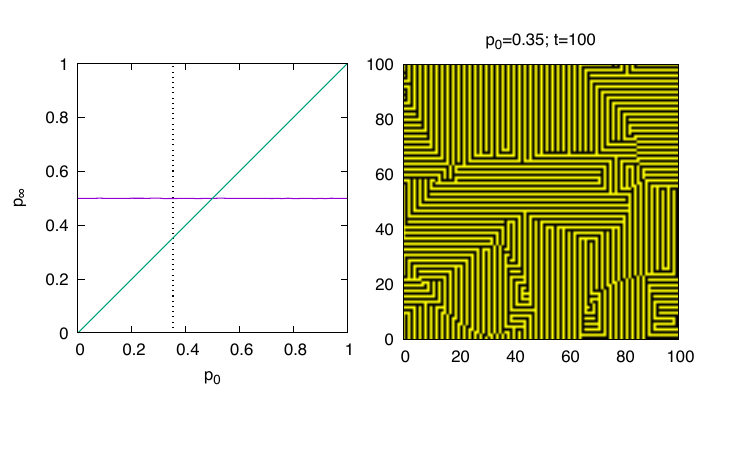}} &
    {\includegraphics[width=0.33\linewidth]{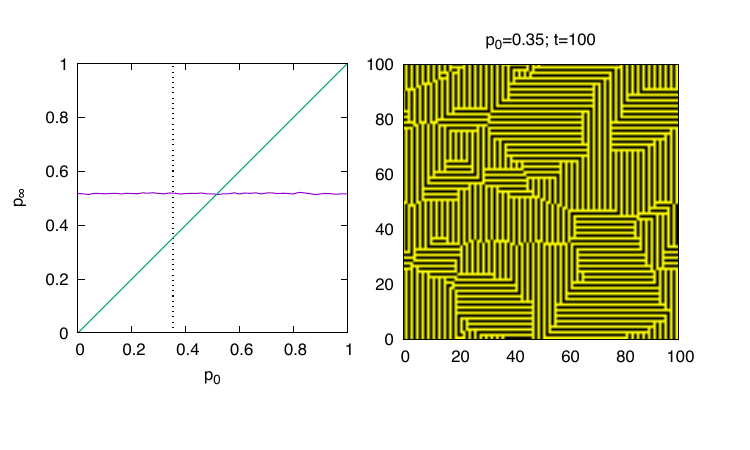}} \\ 
    
     $M\!LE6s:\rho_\infty=0.65$ & $M\!LE7s:\rho_\infty=0.75$  & $M\!LE8s:\rho_\infty=0.85$\\
    {\includegraphics[width=0.33\linewidth]{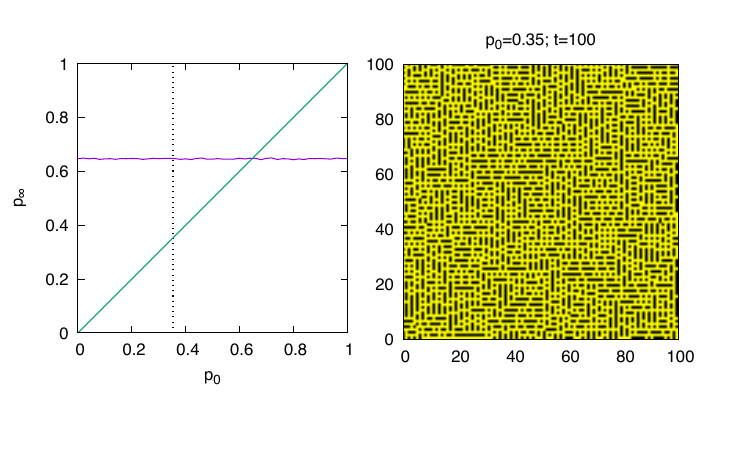}} &
    {\includegraphics[width=0.33\linewidth]{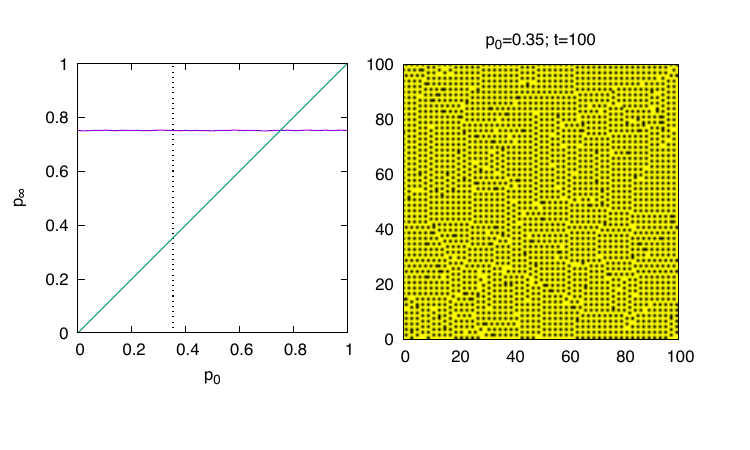}} &
    {\includegraphics[width=0.33\linewidth]{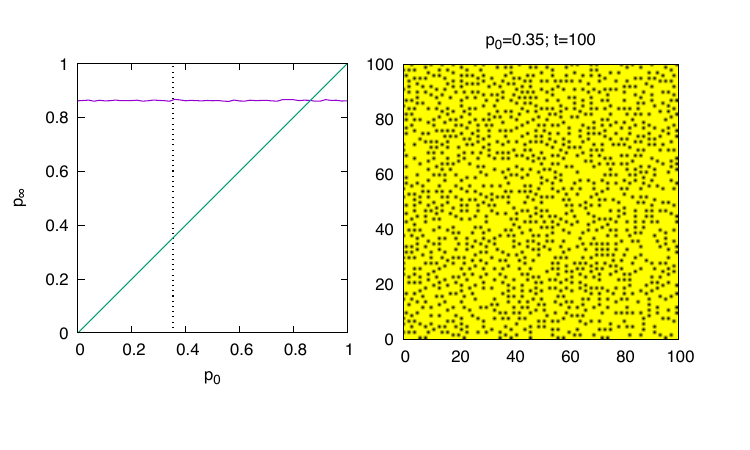}}\\
  \end{tabular}
    \caption{Results for lattices of $100\times100$ sites of fully asynchronous updating of totalistic minority rules $M\!LEXs$, $T=100$.  Left: Final density $\rho_\infty$ vs initial density $\rho_0$. Right:  asymptotic patterns for initial density corresponding to the red dotted line. There is no dependence on the initial density. Except for $X=0$ and $X=8$ (chaotic), all other patterns are composed of static domains, that may finally merge. The pattern for $M\!LE9s$ ($X=9$) is all ones ($\rho_\infty=1$), and the pattern for $M\!L0s$ is all zeros ($\rho_\infty=0$).  }
    \label{fig:LEs}
\end{figure}

A cellular automaton is composed by a set of cells, indexed by a integer $i$, that, at time $t$, can assume one of a given set of states, $s_i(t)$. The cells are connected by a network $a_{ij}\in \{0,1\}$, and the state of the cells connected to a given one (its neighborhood or vicinity) is denoted by $@v_i(t)$:
\eq{
    @v_i(t) = \{s_j(t) : a_{ij}=1\}.
}

The evolution of the state of the cells is given by a transition probability $\tau(s'|@v)$, 
which express the probability that the next state of a cell is $s'$ if the present state of the neighborhood is $@v$. The normalization of the transition probability is 
\eq{
    \sum_{s'} \tau(s'|@v) = 1.
}

This transition probability is generally the same for all cells (homogeneous CA) and is applied simultaneously. 

We consider here  a two-dimensional lattice of size $N\times N$ and  Boolean CA, for which $s_i\in \{0,1\}$. Cells  are sequentially numbered from left to right and from top to bottom, so the index $i$ of a cell of coordinates $(x_i, y_i)$ is given by 
\eq{
    i = y_i N + x_i; \; x_i = i \bmod N; \; y_i = \lfloor i/N\rfloor.
}
All operations on $x_i$ and $y_i$ are modulo $N$.

In our case, the neighborhood of a cell is formed by the 8 nearest cells plus the cell itself. We shall name the  cells in the neighborhood (Fig.~\ref{fig:model}) as 
\begin{gather*}
    C \equiv s_i,\\
    E \equiv s_{i+1},\; 
    W \equiv s_{i-1},\; 
    N \equiv s_{i+N},\; 
    S \equiv s_{i-N},\\
    NE \equiv s_{i+N+1},\;
    NW \equiv s_{i+N-1},\;
    SE \equiv s_{i-N+1},\;
    SW \equiv s_{i-N-1}.\\
\end{gather*}

A \textbf{totalistic} function $f$ is completely symmetric with respect to its arguments and therefore only depends on their sum,  
\eq{
    f(x_1, x_2, x_3, \dots) = g\left(\sum_i x_i\right).
}

In order to use a compact notation, we shall indicate with $H$ the sum of the state of  cells at distance 1 and $\sqrt{2}$,
\eq{
 H = N+S+E+W+NE+NW+SE+SW,
 }
and $M=H+C$. 

A totalistic transition probability depends only on $M$, while an outer totalistic transition probability depends separately on $C$ and $H$.

Clearly, $0\le M \le 9$, and therefore the totalistic transition probability set can be seen as a 9-component vector $P(M)\equiv \tau(1|M)$. The outer totalistic transition probabilities can be seen as two vectors $P_0(H) \equiv \tau(1|C=0, H)$ and $P_1(H)\equiv \tau(1|C=1, H)$.

For a  \emph{deterministic} CA the transition probability vector is formed only by zeros or ones. The evolution function for deterministic CA is also called a ``rule''.

Following Ref.~\cite{Vichniac1984}, we can introduce a series of abbreviations for denoting the rules of two-dimensional deterministic, totalistic or outer totalistic, Boolean CA. For a totalistic rule one uses a $M$ followed by the list of values which give $s'=1$. For instance, the majority rule can be written as $M56789$. For such rules can also say that this rule gives 1 if $M\ge X$ (or $M>Y$; $Y=X-1$), and can be indicated as $MGEX$ (or $MGY$). So the same majority rule can be written as $MGE5$ or $MG4$. Similarly, we can indicate the minority rule as $M\!LE4$ or $M\!L5$, respectively. 

For outer totalistic rules we use the symbol $H$ and we repeat the list for the cases $s=0$ and $s=1$. For instance, the game of Life~\cite{Gardner1970} is  an outer totalistic rule for which $P_0(3)=P_1(2)=P_2(3)=1$ and zero otherwise. It can be denoted as $H3H23$. 

The identity rule, which does not change anything, is again an outer totalistic table with $P_0(H)=0$ and $P_1(H)=1$. It should be indicated as $HHGE0$ but for simplicity it will be denoted by $I$. 

Finally, we append a ``$p$'' to indicate parallel updating, a ``$s$'' for fully asynchronous (``serial'') updating, and a ``$m$'' (``mixed'') otherwise. 

The  patterns for extended majority rules of the type $MGEX$, $X\in \{0,\dots,9\}$ are reported in Fig.~\ref{fig:GE}. The patterns do not depend on the synchronicity of the updating and in general show a dependence on the initial density $\rho_0$ (fraction of one cells). 

A similar behavior is shown by minority rules of the type $M\!LEXp$, with parallel updating, as shown in Fig.~\ref{fig:LEp}. Since in general the local zero-neighborhood is mapped to one, and (except for $M\!LE9$), local one-neighborhood is mapped to zero, the patterns flip at each time step. 
Also, in this case, the patterns and the asymptotic densities depend on the initial density. 

On the contrary, for fully asynchronous minority rules of the type $MLEXs$, the patterns are independent of the initial values, and the final density only depends on the rule used.

The patterns $MLEXs$ are examples of the possible action of an agent with a pure (deterministic) strategy, when the evolution of the world is given by the identity rule (i.e., it keeps all modifications made), since in this case the application of the rule determines the final density regardless of the initial one. 

\section{Learning}\label{sec:learning}

\begin{table}[t]
 \caption{Reinforcement learning of strategy for $s=0$. If the flip of the spin  does not affect $m$ or it overcomes the target, the strategy does not change. Otherwise, if the flip ($s'=1$) brings $m'$ closer than $m$ to the target, the effect of flipping is reinforced ($\Delta P(m)>0$), otherwise it is decreased. The effects on $\Delta P(m)$ for $s=1$ are reversed in sign. Other combinations are not possible.}
    \label{tab:strategy}
    \centering
    \begin{tabular}{c|c|c|c|c}
        \textbf{condition} & $f(\bar{m}, m)$ & $f(\bar{m}, m)$ & $f(m', m)$ & $\Delta P(m)$ \\
        \hline
         $m'=m$ & any & any & 0 & $0$\\
         $m'>\bar{m} > m$ & $1$ & $-1$ & any & $0$\\
         $m'<\bar{m} < m$ & $-1$ & $1$ & any & $0$\\
          \hline
         $\bar{m} > m'>m$ & $1$ & $1$ &$1$ & $+2r$ \\
         $\bar{m} = m'>m$ & $1$ & $0$ &$1$ & $+r$ \\
         $\bar{m} > m>m'$ & $1$ & $1$ &$-1$ & $-2r$ \\
          \hline
         $\bar{m} = m>m'$ & $0$ & $1$ &$-1$ & $-r$ \\
         $\bar{m} = m<m'$ & $0$ & $-1$ &$1$ & $-r$ \\
          \hline
         $\bar{m} < m<m'$ & $-1$ & $-1$ &$1$ & $-2r$ \\
         $\bar{m} = m'<m$ & $-1$ & $0$ &$-1$ & $+r$ \\
         $\bar{m} < m'<m$ & $-1$ & $-1$ &$-1$ & $+2r$ \\
         \end{tabular}
\end{table}

Agents apply  reinforcement learning~\cite{Reinforcement2012} to reach their goal. The idea is the following: the agent has a target density $\bar{m}$; it measures the actual density $m$ in the sensing area (its neighborhood), flips the central spin (the actuator area), and measures the  density $m'$ at the following time step. 

If the flip does not affect the density, or if $m<\bar{m}$ and $m' > \bar{m}$ (or the opposite, $m>\bar{m}$ and $m' < \bar{m}$), nothing happens. 
Otherwise, if the flip brings $m'$ closer than $m$ to the target, the effect of flipping is reinforced, in the sense that $P(m)$ is modified in order to increase the probability of the effect of flipping, otherwise it is decreased.  This strategy is summarized in Table~\ref{tab:strategy}, columns 1 and 5.

We can make the evolution of strategy more compact. Let us define a test function
\[
    f(x,y) =\begin{cases}
       -1 & \text{if $x<y$},\\
         0 & \text{if $x=y$},\\
        1 & \text{if $x>y$}.\\
    \end{cases}
\]

The variation of the strategy is thus given by 
\[
    \Delta P(m) = \left[f(\bar{m},m) + f(\bar{m},m')\right]\cdot f(m', m)\cdot (1-2s) r,
\]
as reported in Table~\ref{tab:strategy}, columns 2-5. 
Therefore, 
\[
 P'(m) = \begin{cases} 
        0 & \text{if $P'(m) + \Delta P(m) \le 0$}\\
        1 & \text{if $P'(m) + \Delta P(m) \ge 1$}\\
        P(m) + \Delta P(m) & \text{otherwise}
    \end{cases}
\]
since $P(m)$ is bounded between 0 and 1. Agents train themselves for a certain number $T$ of epochs.

\section{Passive environment}\label{sec:identity}

Let us first investigate what happens when the environment is passive, i.e., it follows  the identity rule. 
We assume here that all agents have the same target, and that there is no interference (other agents whose target areas are inside the sensing area of the one under investigation). In this case the evolution of the strategy is quite simple. 

For $s=0$, given a measurement $m<\bar{m}$, we get $m'=m+1$ so $P(m)$ increases by a positive amount, while it decreases if $m>\bar{m}$, and vice versa for $s=1$.

\begin{figure}
    \centering
    \begin{tabular}{cc}
        (a) & (b) \\
        \includegraphics[width=0.4\linewidth]{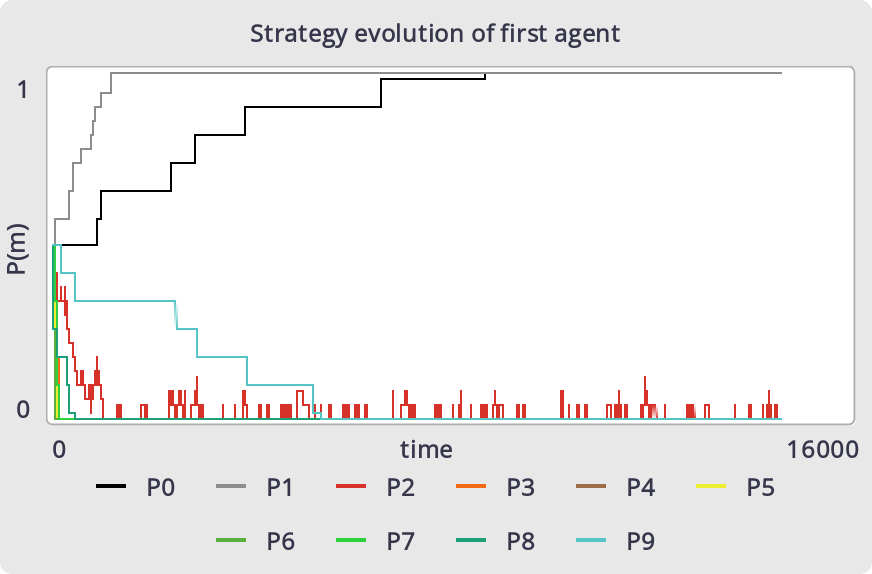}
        \includegraphics[width=0.4\linewidth]{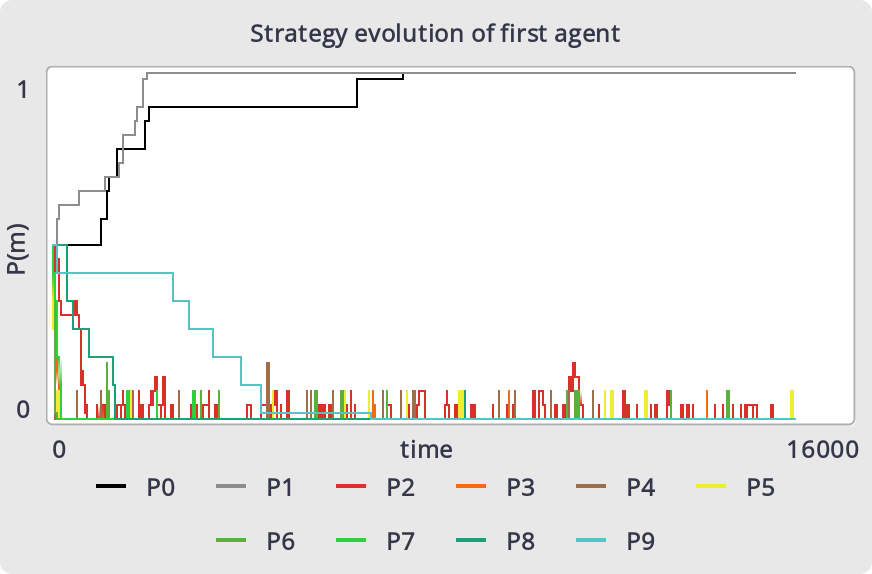}
    \end{tabular}
    \caption{The evolution (learning) of the strategy $P(m,t)$ vs time for the identity rule, with $\bar{m}=2$ and $N=13$. (a) One single agent; (b) 10 agents.}
    \label{fig:evolution1}
\end{figure}

If the agent is able to sample all possible values of $m$ for a sufficient large number of times, the final strategy is just $P(m)=1$ for $m<\bar{m}$, and $P(m)=0$ for $m>\bar{m}$, i.e., a minority rule $MLEs$ for a single agent, or $MLEm$ for more than one agents. As seen in Section~\ref{sec:TCA}, this is a pure strategy that gives an asymptotic density independent on the initial one (for the passive environment), and thus can approximate the target $\bar{m}$. 

It is evident that here the learning rate $r$ plays little role, while  the ``variability'' of the environment is important: notice the slow learning of $P(0)$ and $P(9)$.

The agent alters the environment while testing, and therefore the density tends to be skewed towards $0.5$, while, when applying the learned strategy, the density tends to that of the fully asynchronous minority rule strategy (Sect.~\ref{sec:TCA}).

The variability of the sampled values of $m$ is improved if more agents are learning at the same time, or by mixing agents with different goals. 

\section{More complex environments}\label{sec:complex}
\begin{figure}
    \centering
    \begin{tabular}{cc}

    \includegraphics[width=0.4\linewidth]{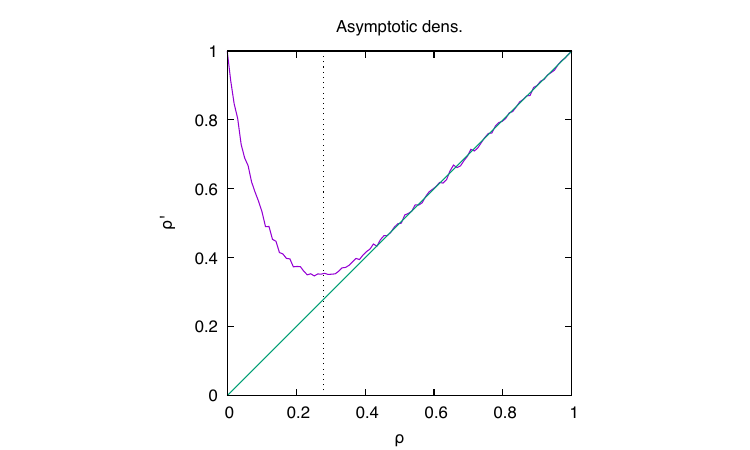} & 
    \includegraphics[width=0.4\linewidth]{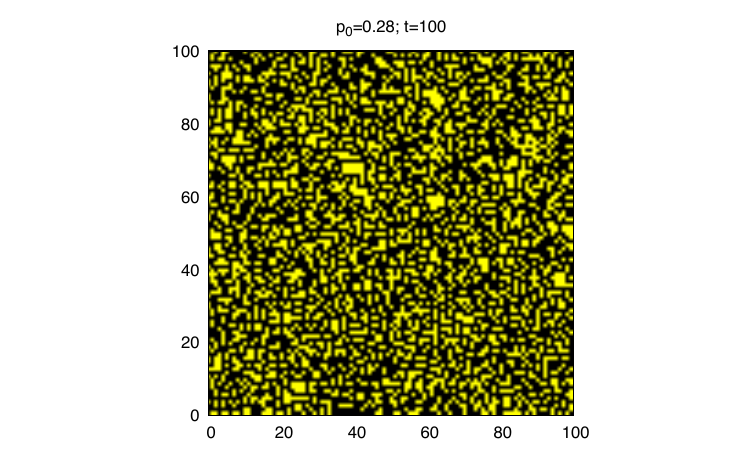} \\
    \includegraphics[width=0.4\linewidth]{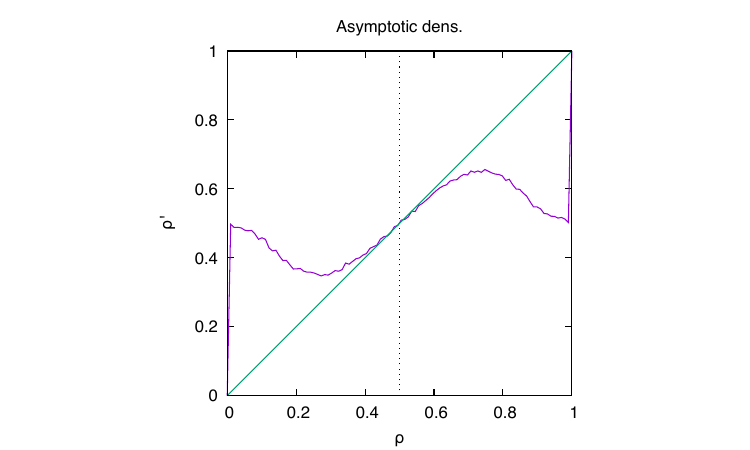} & 
    \includegraphics[width=0.4\linewidth]{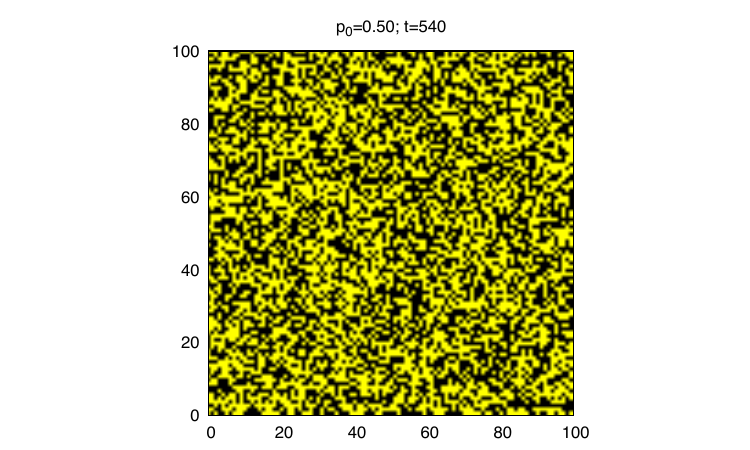} \\
    
    \includegraphics[width=0.4\linewidth]{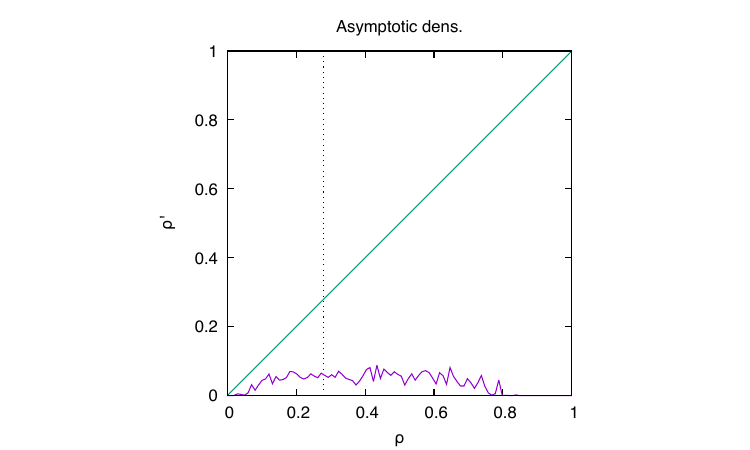} & 
    \includegraphics[width=0.4\linewidth]{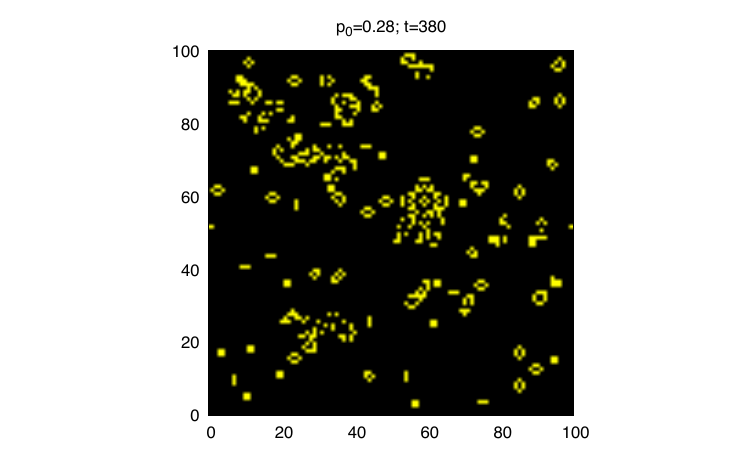} \\
    \end{tabular}
    \caption{Results for lattices of $100\times100$ sites of fully synchronous updating of outer totalistic minority rules,  $T=100$. From top to bottom: $H0HGE1p$, $H08H1234567p$ and Life $H3H23p$. Left: the asymptotic density as a function of the initial one. Right: a snapshot of a configuration corresponding to the initial density marked by the red dotted line in the left column.}
    \label{fig:complex}
\end{figure}

In this preliminary paper we explore the consequences of three rules for the evolution of the environment: the frustrated identity $H0HGE1p$, i.e, the identity except that for $H=0$ $C=0$ gives $C'=1$ and $C=1$ gives $C'=0$; the double frustrated identity $H0HGE1p$, which, ``flips'' also the case $H=8$, and the Game of Life, $H3H23p$. As shown in Figure~\ref{fig:complex}, the correspondence between the initial and the final density (which is fully diagonal for the identity rule), shows more and more ``unnatural'' targets: for $H0HGE1p$ the target $\bar{\rho} < 0.38$ is not in the natural range, for $H0HGE1p$ also the target $\bar{\rho} > 0.62$ is unnatural, for Life $H3H23p$ only targets $\bar{\rho} < 0.03$ are natural.

\begin{figure}
    \centering
    \begin{tabular}{cc}
    \includegraphics[width=0.4\linewidth]{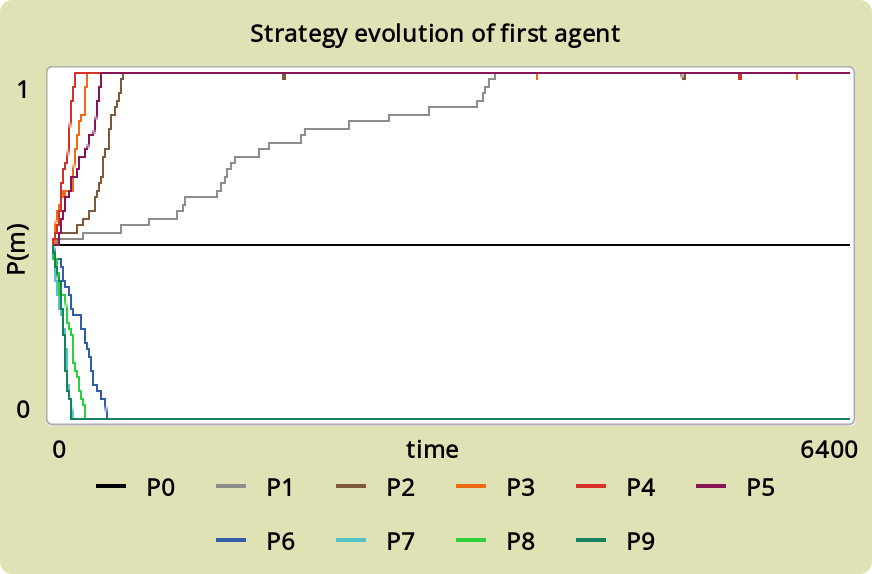} &
    \includegraphics[width=0.4\linewidth]{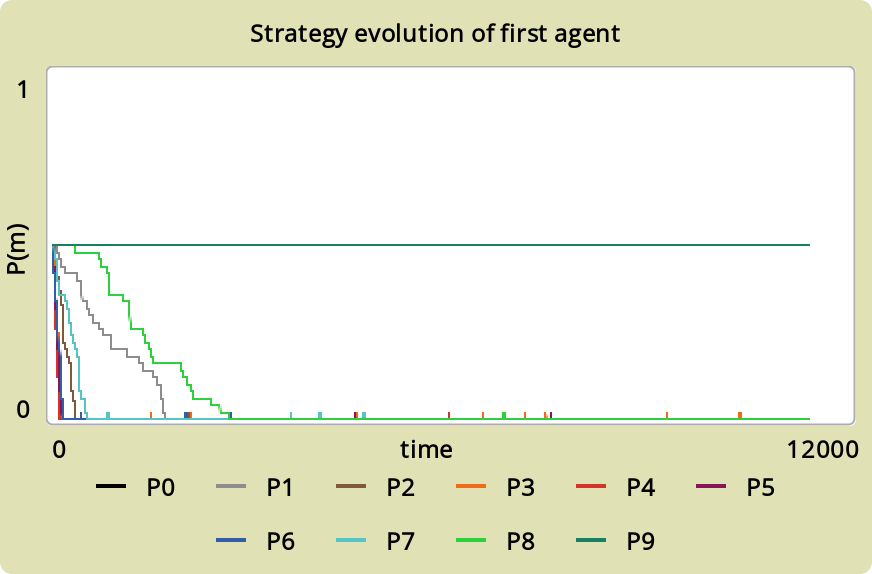} 
    \end{tabular}
    \caption{The learning process for the $H0HGE1p$ environment. Left: $\bar{m}=5.2$ ($\bar{\rho}=0.6$). Right: $\bar{m}=0$ ($\bar{\rho}=0$). }
    \label{fig:LearnedH0HGE1}
\end{figure}

For the frustrated majority rule, a target within the natural range (say, $\bar{m}=5.4$ -- $\bar{\rho}=0.6$) gives the same results as in the passive environment, since it involves the "identity" portion of the density graph. However, agents never learn what to do for local densities that are forbidden by the rule, for instance for $m=0$ in the $H0HGE1p$ environment, $P(0)$ always remains at the original value $0.5$ since agents can never be successful in changing such a local configuration (Fig.~\ref{fig:LearnedH0HGE1}-left), the same happens also for $P(9)$ and environment $H08H1234567p$. 

Setting a target outside the natural area, for instance $\bar{m}=0$ for $H0HGE1p$, correctly induces all ``evolving'' probabilities (in this case, all except $P(0)$) to reach the ``right'' values (Fig.~\ref{fig:LearnedH0HGE1}-right), but the application of the rule can only partially alter the local density. In the last case, the density obtained by applying the rule with 20 agents on a $100\times100$ lattice only gives $\rho_\infty\simeq 0.3$. 

\begin{figure}
    \centering
    \includegraphics[width=0.4\linewidth]{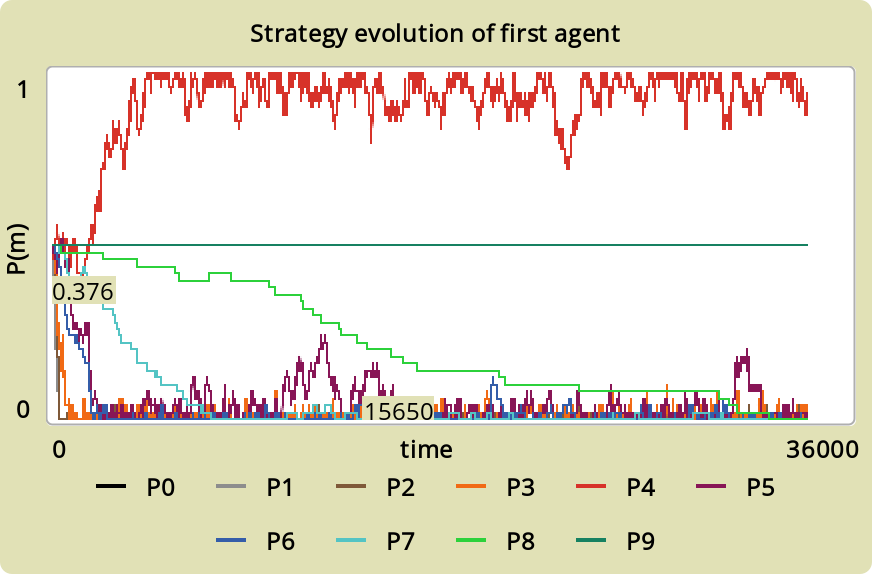}
    \caption{The learning process for the Game of Life, $H3H23p$ environment, and $\bar{m}=0.03$.}
    \label{fig:LearnedH3H23}
\end{figure}

For the Game of Life,  $H3H23p$, let us set a  target in the natural range, $\bar{\rho}=0.01-0.03$  ($0<\bar{m}<0.27$), but different from the trivial ones $\bar{\rho}=0$ (extinction). The first result is that a single agent cannot fulfill the task, since  it always brings the system to extinction (the asymptotic ``animals'' in life~\cite{Bagnoli1991} are quite sensible to perturbations), while the presence of more agents is able to ``keep alive'' the environment. However, agents are unable to learn what to do for local measurements with $m=0$, since all actions that they perform on such a local configuration give $C'=0$ (no improvement towards the target),  so their are left with $P(0)=0.5$ and the learned rule for small values of $m$ is $P(1)=P(2)=P(3)=0$ (only $P(4)=1$), so the application of such a rule always brings the system to extinction, Figure~\ref{fig:LearnedH3H23}. 

Setting a higher target, like $\bar{m}=1$, brings $P(0)=1$, but this is still insufficient to get an asymptotic density different from zero. Only setting unreachable targets like $\bar{m}=9$ induces all $P(m)=1$, with a final density only slightly above the ``natural'' one, $\bar{\rho}=0.06$ with 20 agents in a $100\times100$ lattice.

\section{Conclusions}\label{sec:conclusion}
We have investigated the problem of moving cognitive agents (modeled as totalistic probabilistic cellular automata) that learn how to modify their environment (modeled as a deterministic, outer-totalistic cellular automata). Agent can sense the local configuration of the environment and alter the center cell, trying to make the local density to go towards their predefined target. We have shown that agents may learn how to approach their goal when the environment is passive, and only partially or not at all if the environment follows an active dynamics.

There are many other aspects of the model to be investigated in future works. 
\printbibliography

\end{document}